\documentclass{aa}
\usepackage[varg]{txfonts}
\usepackage{graphicx}
\usepackage{amssymb}
\usepackage{natbib,twoopt}
\usepackage[breaklinks=true]{hyperref} 
\bibpunct{(}{)}{;}{a}{}{,} 
\usepackage{upgreek}
\usepackage{mathptmx}
\usepackage{booktabs,siunitx}
\usepackage{mathtools}
\usepackage{dcolumn}
\usepackage{amsmath}
\usepackage{color}

\begin{document}

\title{The glitch activity of neutron stars}

\author{J.R. Fuentes \inst{\ref{uc}}
\and C. M. Espinoza\inst{\ref{usach}}
\and A. Reisenegger\inst{\ref{uc}}
\and B. Shaw\inst{\ref{man}}
\and B. W. Stappers\inst{\ref{man}}
\and A. G. Lyne\inst{\ref{man}}}

\institute{Instituto de Astrof\'isica, Pontificia Universidad Cat\'olica de Chile, Av. Vicu\~na Mackenna 4860, 7820436 Macul, Santiago, Chile \label{uc} \and Departamento de F\'isica, Universidad de Santiago de Chile, Avenida Ecuador 3493, 9170124 Estaci\'on Central, Santiago, Chile \label{usach} \and Jodrell Bank Centre for Astrophysics, School of Physics and Astronomy, The University of Manchester, Manchester M13 9PL, UK\label{man}}

\date{Accepted XXX / Received YYY}

\abstract{We present a statistical study of the glitch population and the behaviour of the glitch activity across the known population of neutron stars. An unbiased glitch database was put together based on systematic searches of radio timing data of 898 rotation-powered pulsars obtained with the Jodrell Bank and Parkes observatories. Glitches identified in similar searches of 5 magnetars were also included. The database contains 384 glitches found in the rotation of 141 of these neutron stars. We confirm that the glitch size distribution is at least bimodal, with one sharp peak at approximately $20\, \rm{\mu\,Hz}$, which we call large glitches, and a broader distribution of smaller glitches. We also explored how the glitch activity $\dot{\nu}_{\rm{g}}$, defined as the mean frequency increment per unit of time due to glitches, correlates with the spin frequency $\nu$, spin-down rate $|\dot{\nu}|$, and various combinations of these, such as energy loss rate, magnetic field, and spin-down age. It is found that the activity is insensitive to the magnetic field and that it correlates strongly with the energy loss rate, though magnetars deviate from the trend defined by the rotation-powered pulsars. However, we find that a constant ratio $\dot\nu_{\rm{g}}/|\dot\nu| = 0.010 \pm 0.001$ is consistent with the behaviour of all rotation-powered pulsars and magnetars. This relation is dominated by large glitches, which occur at a rate directly proportional to $|\dot{\nu}|$. For low $|\dot\nu|$, only small glitches have been detected, making the inferred glitch activity formally lower than that predicted by the constant ratio, in many cases zero. However, we can attribute this to the low predicted rate for large glitches, together with the insufficient observing time, which makes it unlikely to detect any large glitches in this range. Taking this into consideration, we show that the behaviour of each rotation-powered pulsar and magnetar is statistically consistent with the above relationship, including those objects where no glitches have been detected so far. The only exception are the rotation-powered pulsars with the highest values of $|\dot{\nu}|$, such as the Crab pulsar and PSR B0540$-$69, which exhibit a much smaller glitch activity, intrinsically different from each other and from the rest of the population. The activity due to small glitches also shows an increasing trend with $|\dot\nu|$, but this relation is biased by selection effects.}


\maketitle

\section{Introduction}

Rotation-powered pulsars (hereafter just "pulsars") and magnetars are believed to be highly magnetized, rotating neutron stars. Their spin can be tracked with high accuracy over years using standard timing techniques \citep{1991ApJ...371..739R,1991ApJ...380..557R,1994ApJ...428..713K}. Their rotation is generally stable and shows a regular spin-down trend (frequency derivative $\dot{\nu}<0$). Nevertheless, many pulsars exhibit sudden increases in their rotation frequency $\nu$, known as glitches. Glitches have relative sizes $\Delta \nu/\nu \sim 10^{-11} - 10^{-5}$, and in most cases, are followed by an increase in the spin-down rate $\dot{\nu}$ of the star, with relative magnitudes $\Delta\dot{\nu}/\dot{\nu} \sim 10^{-5}-10^{-2}$. Several mechanisms have been suggested to explain these phenomena \citep[see, for instance,][for a recent review of glitch models]{2015IJMPD..2430008H}. The lack of observed radiative changes associated with these events in nearly all rotation-powered pulsars suggests an internal origin. In this context, glitches are believed to be caused by a rapid transfer of angular momentum from a neutron superfluid in the inner crust to the rest of the star \citep{1975Natur.256...25A,1984ApJ...276..325A}. However, glitches in magnetars (and also in the high magnetic field pulsars J1119$-$6127 and J1846$-$0258) are sometimes accompanied by radiative changes, and could have a different origin \citep{2014ApJ...784...37D,2017ARA&A..55..261K}.\\

Observations of glitches represent one of the very few instances through which we can indirectly inspect the interior of a neutron star. For example, attempts have been made to use glitches to put constraints on the structural properties of neutron stars, their masses, and the equation of state of dense matter \citep{1999PhRvL..83.3362L,2012PhRvC..85c5801C,2013PhRvL.110a1101C,2015SciA....1E0578H,2017NatAs...1E.134P}. Recent studies of the rotation of young neutron stars, such as the Vela and Crab pulsars, suggest that the effects of glitches over the long-term spin evolution are comparable to the effects driven by magnetic braking or stellar winds \citep{1996Natur.381..497L,2015MNRAS.446..857L,2017MNRAS.466..147E}.\\

Trends in the glitch behavior of pulsars have been slowly emerging as the number of detected glitches increases. \citet{1990Natur.343..349M} and \cite{2000MNRAS.315..534L} showed that the glitch activity describes a linear behavior with the spin-down rate $|\dot{\nu}|$ of the star. \citet{2011MNRAS.414.1679E}, using a larger sample, confirmed this result and found that the glitch activity decreases rapidly in pulsars with low spin-down rate. They also established that young pulsars exhibit glitches more often than old pulsars, and found that the size distribution of all observed glitches has a bimodal appearance. Furthermore, they found that most of the largest glitches are produced by a small group of relatively young pulsars, suggesting that the bimodality of the glitch size distribution could be produced by two different classes of pulsars or by a glitch mechanism that evolves as pulsars age.\\

This work is intended to be an extension of the previous study by \citet{2011MNRAS.414.1679E}. It presents the building and analysis of a sample of glitches that is unbiased towards their presence in the data. The events were taken from published systematic records of timing observations of hundreds of pulsars. We focus on the frequency step sizes and their rate (the \emph{glitch activity}) and study how they depend on long-term spin properties (spin frequency, spin-down rate, and  combinations of these, such as energy-loss rate, magnetic field, and spin-down age). This paper is organised as follows. Section \ref{newdat} describes the new database and how neutron stars and glitch detections were selected to avoid bias in our sample. In \S \ref{s2} we analyze the glitch size distribution and classify glitches according to their sizes. Section \ref{methodglitch} presents a study of the cumulative effect of glitches on the rotation of neutron stars and a discussion of the relation between the glitch activity and the spin-down rate. Finally, \S \ref{disc} and \S \ref{summary} show the Discussions and Conclusions of the paper, respectively.

\section{The new database} \label{newdat}
Today, more than 700 pulsars are regularly monitored at the Jodrell Bank Observatory (JBO), some of them from as early as 1978 \citep{2004MNRAS.353.1311H}. These long-term observations are essential to finding glitches and studying their properties. In order to build a sample that is as unbiased as possible, we included all pulsars that have been regularly monitored for glitches in clearly defined time spans, regardless of whether glitches were found or not. Selecting only those pulsars for which glitches have been detected would bias the sample towards the presence of glitches. According to this scheme, we included 778 pulsars monitored at JBO, containing 296 glitches in the rotation of 111 pulsars. These glitches are the JBO events in \cite{2011MNRAS.414.1679E} plus 69 newer glitches measured until 2015 and published in the JBO online glitch catalog \footnote{\url{http://www.jb.man.ac.uk/pulsar/glitches.html}} (Shaw et al. in preparation).\\

In order to expand the sample, we also included Parkes observations of 118 pulsars, which show 73 glitches in the rotation of 23 pulsars, as reported by \cite{2013MNRAS.429..688Y}. In case of overlap between the observation spans of the JBO and Parkes pulsars, we considered the earliest and the latest epoch between the two to define the start and end of the searched time spans.\\

In order to improve the statistics for pulsars with small characteristic ages ($\uptau_{\rm{c}} = -\nu/2\dot{\nu}$), we also added the two X-ray pulsars PSRs J1846$-$0258 and B0540$-$69, which have been monitored for about 15 years each and have been searched for glitches \citep{2015ApJ...812...95F,2011ApJ...730...66L}. With these two additions, the database contains all the rotation-powered pulsars known with $\uptau_{\rm{c}} < 2\, \rm{kyr}$. It is not possible to obtain a complete sample for pulsars of larger characteristic ages because many of them have either not been regularly monitored or not been searched for glitches.\\

Finally, to compare the glitch activity between rotation-powered pulsars and magnetars, we included the observations of five magnetars. They have been observed continuously for 16 years on average, and \citet{2014ApJ...784...37D} reported a set of 11 glitches in the whole of their timing dataset.\\

We constructed a database containing rotational information ($\nu$, $\dot{\nu}$), with $\dot{\nu}$ corrected for the Shklovskii effect \citep{1994ApJ...421L..15C}, glitch measurements $\Delta \nu$, and the observation spans over which glitch searches have been performed (on average, $17.5$ $\rm{years}$ for each pulsar). Here, $\Delta \nu$ corresponds to the frequency increase due to the glitch. We did not take $\Delta \dot{\nu}$ steps or recoveries into account because not all glitches have these parameters measured in a consistent way.\\

\begin{figure}[h!]
\centering
\includegraphics[width=90mm]{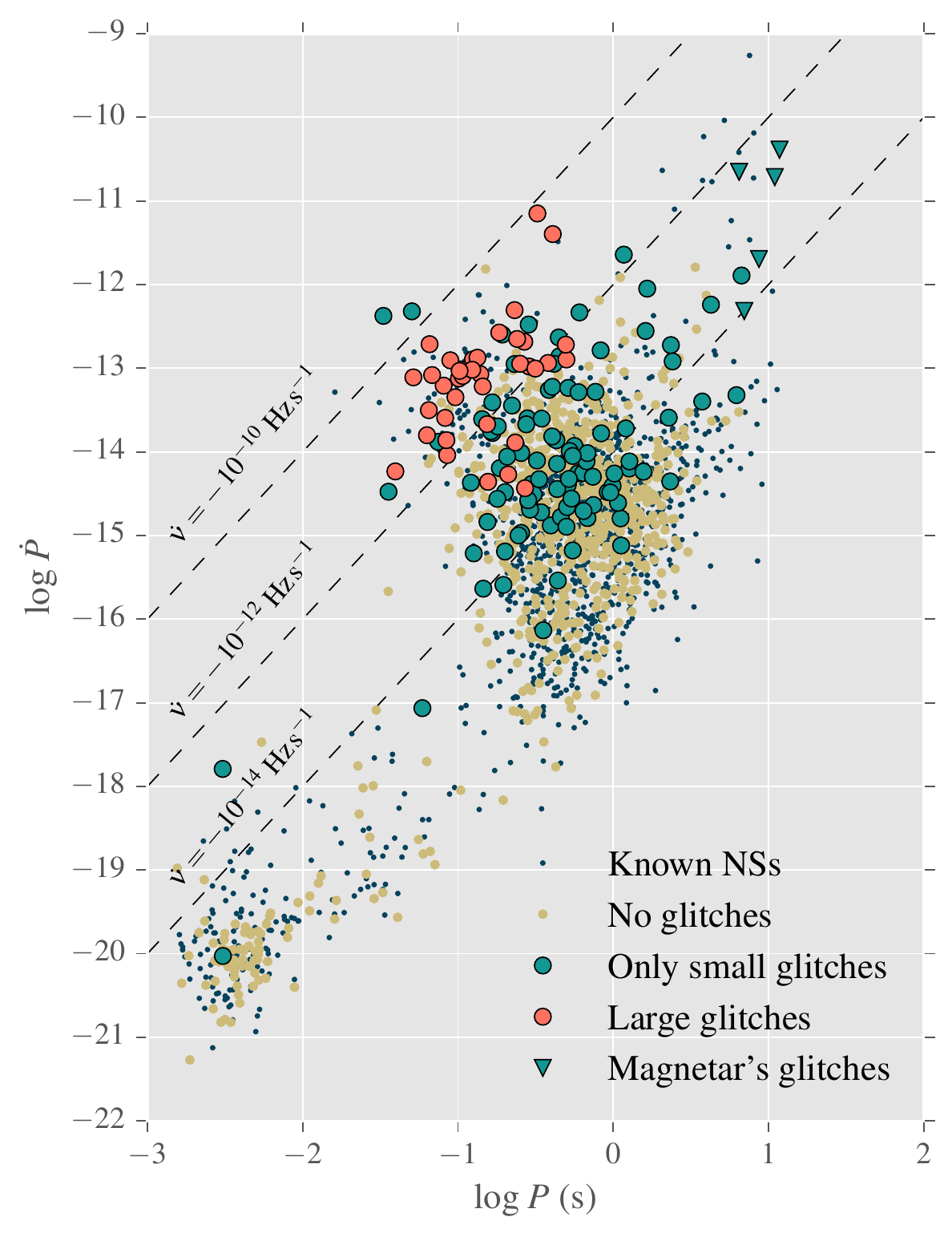}
\caption[Rotation period versus its time derivative ("$P$-$\dot{P}$ diagram") for all known neutron stars. Lines of constant spin-down rate $\dot{\nu}$ are shown and labelled. The small dark blue and medium light amber dots denote the known neutron stars not in our database, and the neutron stars in our database with no detected glitches, respectively. The large orange and turquoise dots represent those pulsars in our database with large glitches, and only small glitches detected, respectively. The turquoise triangles correspond to the magnetars in our database with only small glitches detected. $P$ and $\dot{P}$ for neutron stars not in our database were taken from the ATNF pulsar catalogue \citep{2005AJ....129.1993M}.]{Rotation period versus its time derivative ("$P$-$\dot{P}$ diagram") for all known neutron stars. Lines of constant spin-down rate $\dot{\nu}$ are shown and labeled. The small dark blue and medium light amber dots denote the known neutron stars not in our database, and the neutron stars in our database with no detected glitches, respectively. The large orange and turquoise dots represent those pulsars in our database with large glitches, and only small glitches detected, respectively. The turquoise triangles correspond to the magnetars in our database, which only have small glitches detected. $P$ and $\dot{P}$ for neutron stars not in our database were taken from the ATNF pulsar catalog \footnotemark  \citep{2005AJ....129.1993M}.} \label{fig1}
\end{figure}

Altogether, our sample contains the rotational information of 903 neutron stars, as shown in Figure \ref{fig1}, with a total of 384 glitches in 141 of them. The sample does not have a well-defined selection criteria, being mostly determined by having pulsars bright enough that they could be regularly monitored without an extreme commitment of observing time. In addition, the observing time spans are not uniform, as additional pulsars were added as they were discovered. On the other hand, it is important to note that the sample includes nearly a third of all pulsars known to date, with representatives across the $P-\dot P$ diagram (see Fig. \ref{fig1}), and none of the pulsars were selected directly because of their glitch properties (presence or absence of glitches, their frequency, or size). Thus, it should be close to the best possible available sample for the study performed in the present paper, and the biases present should not affect our conclusions.
\footnotetext{\url{http://www.atnf.csiro.au/research/pulsar/psrcat}} 

\section{The glitch size distribution} \label{s2}
\begin{figure}[h!]
\centering
\includegraphics[width=90mm]{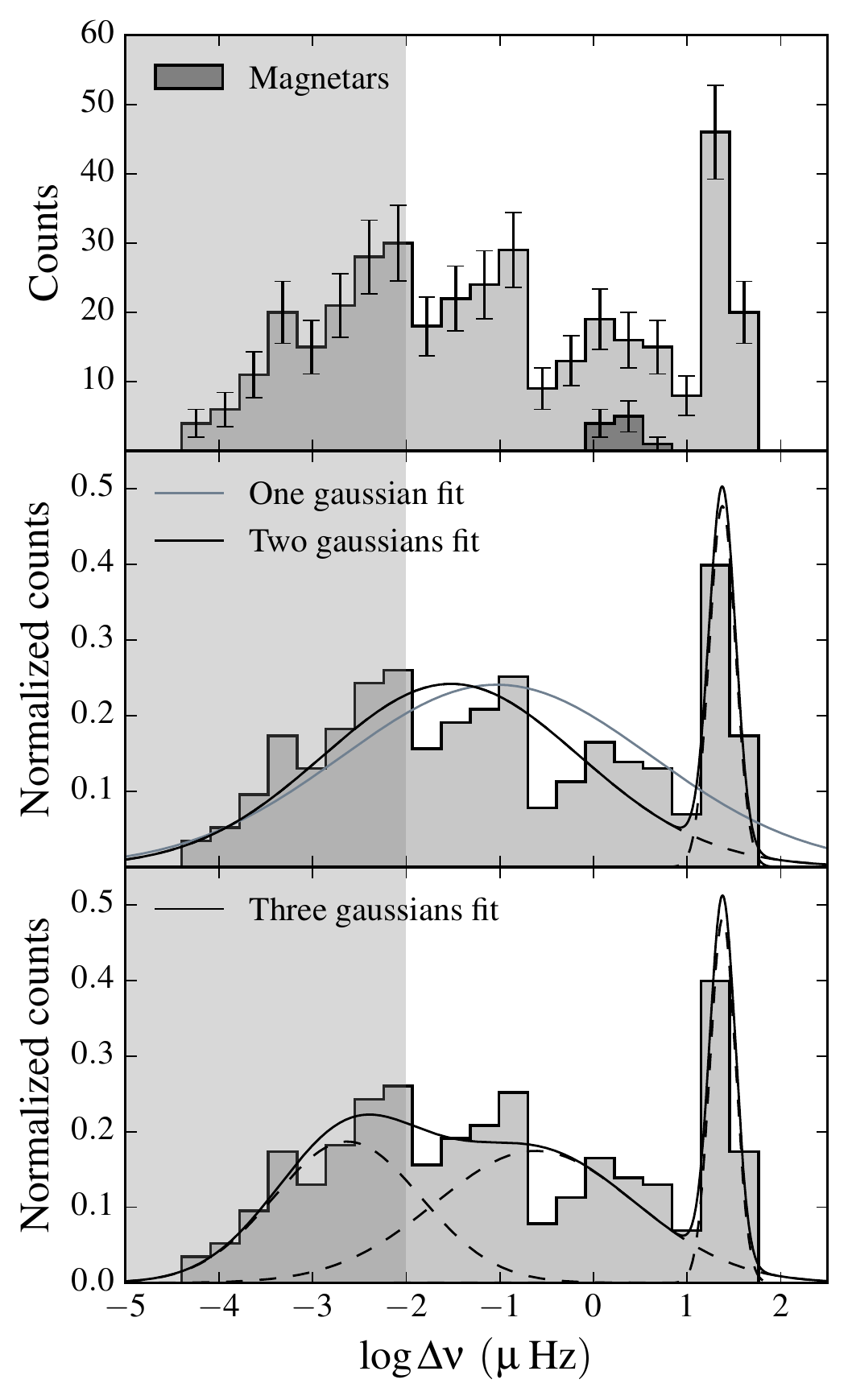}
\caption{Histogram of the glitch size $\Delta \nu$ of all glitches in our database. In the upper panel the error bars correspond to the square root of the number of events per bin. The middle and lower panels show the best fits with one, two, and three Gaussians. Magnetar glitches were not included in the latter panels nor in the fits. The solid and dashed lines represent the best fits and their components, respectively. The shaded region indicates that glitches of sizes smaller than $0.01$ $\rm{\mu\, Hz}$ may be missing due to detectability issues.} \label{fig2}
\end{figure}

The distribution of the glitch magnitude $\Delta\nu$ of all glitches in our database is shown in Figure \ref{fig2}, and is in agreement with the bimodal shape reported by \citet{2011MNRAS.414.1679E} and \citet{2017PhRvD..96f3004A}. There is a broad distribution of small glitches and a very narrow distribution of large glitches which peaks around $20\, \rm{\mu\, Hz}$. It is worth mentioning that this peak contains 70 large glitches detected in 38 different pulsars, where the main contributor is PSR J1420$-$6048, with 5 large glitches. This confirms that this peak of large glitches is by no means the effect of only a handful of pulsars.\\

As noted by \citet{2011MNRAS.414.1679E}, the left edge of the distribution is unconstrained since detections of very small glitches are strongly limited by the cadence of the observations, their sensitivity, and the intrinsic rotational noise of the pulsar. Given that these properties vary from pulsar to pulsar, it is not straightforward to set a universal lower detectability limit for the glitch size distribution of all pulsars (Fig. \ref{fig2}). However, based on the comparison of glitch sizes obtained by different authors, \citet{2011MNRAS.414.1679E} argued  that for sizes below $\Delta \nu/\nu \sim 10^{-8}$ the sample is likely to start being incomplete. Using the detectability limits for individual sources proposed by \citet{2014MNRAS.440.2755E}, it is possible to calculate average detection limits for all glitches.  For an observing cadence of 30 days and a rotational noise (or minimum sensitivity) of 0.01 rotational phases, glitch detection is severely compromised below sizes $\Delta\nu \sim 10^{-2}\, \rm{\mu Hz}$. This is consistent with the $\Delta\nu/\nu$ figure quoted above (for young pulsars). Hence glitches with magnitudes below this limit are likely to be missed, especially if their frequency derivative steps are larger than $|\Delta\dot{\nu}|\sim 10^{-15} \, \rm{Hz\, s^{-1}}$ \citep[see][]{2015aska.confE..43W}. This is a conservative limit because many pulsars are observed more frequently and exhibit lower noise levels.\\

In the following, we model the glitch size distribution as a sum of Gaussian distributions. We note, however, that the detection issues discussed above must be considered when using such functions to describe the population of small glitches. Models composed of one and up to four Gaussians were tested against the data. The best fits obtained are shown in Figure \ref{fig2} (middle and lower panels; see Table \ref{akairesults} for the parameters of the fits). Using Akaike's information criterion (hereafter AIC, see Appendix \ref{akai} for a brief description of the test and the results obtained), we conclude that among the whole set of candidate models, a mixture of three Gaussians gives the best description of the glitch size distribution. All fits performed with more than one Gaussian give a common component (with nearly identical parameters) that contains all the glitches with magnitudes $\Delta\nu \geq 10\, \rm{\mu\, Hz}$. Because of this and the multimodal nature of the distribution, we classify glitches as large and small using $\Delta\nu = 10\, \rm{\mu\,Hz}$ as the dividing line. We also made fits for the $\Delta \nu/\nu$ distribution, and the AIC suggests, as for $\Delta\nu$, a non-unimodal behavior, although with two broader peaks of similar height. This further supports the multimodal interpretation of the glitch size distribution.

\section{Glitch activity} \label{methodglitch}
A practical way to quantify the cumulative effects of a collection of spin-ups due to glitches on a pulsar's rotation is through the glitch activity. This parameter is defined as the time-averaged change of the rotation frequency due to glitches. Due to the short length of the observation spans available, it is not possible to detect enough glitches for a robust estimation for each pulsar.  To avoid this problem, we studied the combined glitch activity for groups of pulsars sharing a common property. Following \citet{2000MNRAS.315..534L} and \citet{2011MNRAS.414.1679E}, the average glitch activity for each group is
\begin{align}
\dot{\nu}_{\rm{g}} = \frac{\sum_i\sum_j \Delta \nu_{ij}}{\sum_i T_i} \text{,} \label{gac}
\end{align}
where the double sum runs over every change in frequency $\Delta \nu_{ij}$ due to the glitch $j$ of the pulsar $i$, and $T_i$ is the time over which pulsar $i$ has been searched for glitches. This analysis includes those pulsars that have been searched, but not found to glitch so far. Since the errors in the measurements of the glitch sizes are smaller than the Poisson fluctuations in the number of glitches detected due to finite observation spans, the errors for $\dot{\nu}_{\rm{g}}$ are estimated as $\delta \dot{\nu}_{\rm{g}} = \sqrt{\sum_i \sum_j \Delta \nu_{ij}^2}/\sum_i T_i$ (for more details see Appendix \ref{errors}). Unlike those presented in \citet{2000MNRAS.315..534L} and \citet{2011MNRAS.414.1679E}, these errors account for the presence of glitches of different sizes. However, our formula still does not take into account the possible contribution of rare, large glitches that were not detected because of the finite monitoring times. This implies that the error bars for $\dot\nu_{\rm{g}}$ are likely underestimated.\\

\subsection{Dependence on spin parameters}
\begin{figure}
\begin{center}
\includegraphics[width=70mm]{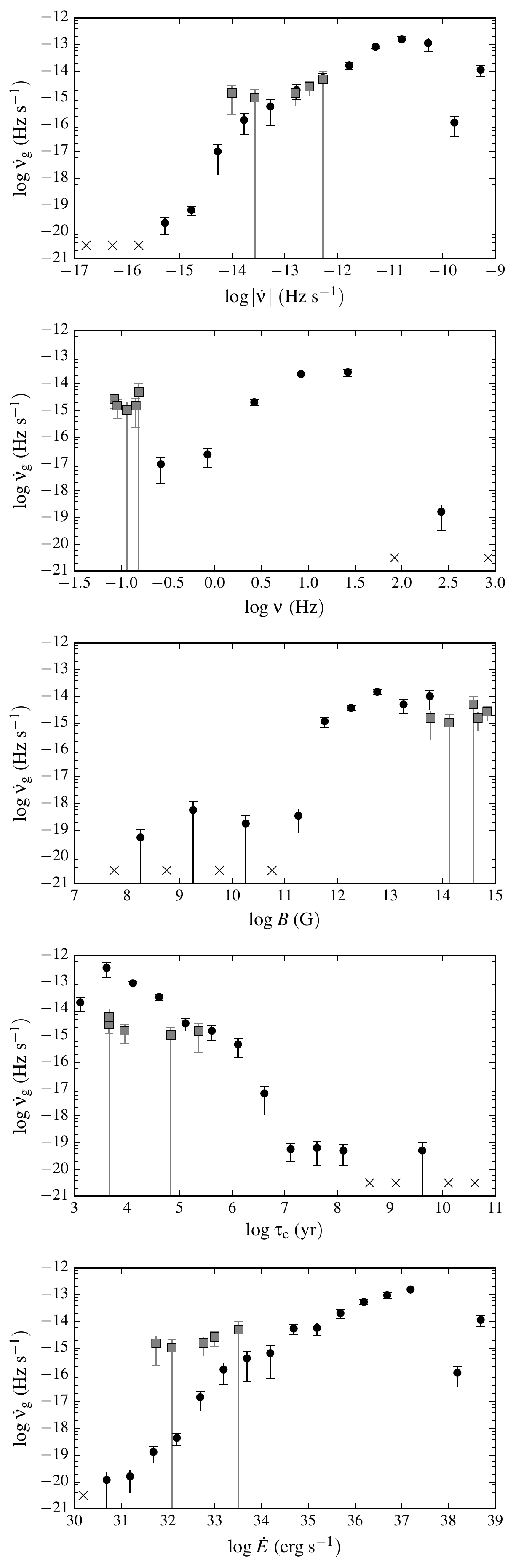}
\caption{Glitch activity as a function of various pulsar parameters (all of them combinations of frequency $\nu$ and its time-derivative $\dot\nu$). Black dots are the $\dot{\nu}_{\rm{g}}$ values calculated according to Eq \ref{gac}, for rotation-powered pulsars grouped in bins of width 0.5 in the logarithm (base 10) of the variable on the horizontal axis. The crosses denote bins (groups of pulsars) with no detected glitches, and the gray squares represent individual magnetars.}
\label{fig3}
\end{center}
\end{figure}

We grouped the pulsars in bins of width equal to $0.5$ in logarithmic scale according to different properties (spin frequency $\nu$, spin-down rate $\dot{\nu}$, and various combinations of these, such as energy-loss rate $\dot{E}_{\rm{rot}} \propto \nu \dot{\nu}$, magnetic field $B \propto \sqrt{\dot{\nu}/\nu^3}$, and spin-down age $\uptau_{\rm{c}} \propto \nu/\dot{\nu}$). Figure \ref{fig3} shows how $\dot{\nu}_{\rm{g}}$ depends on these variables. Considering only rotation-powered pulsars, we observe that $|\dot{\nu}|$, $\dot{E}_{\rm{rot}}$, and $\uptau_{\rm{c}}$ appear to give good correlations, whereas there are no clear correlations with $B$ and $\nu$.\\

It is also apparent in Figure \ref{fig3} that the glitch activity of the magnetars with the smallest characteristic ages is lower than that of the rotation-powered pulsars with similar characteristic ages. However, their activity is larger than that of pulsars of equal spin-down power. The only parameter for which the glitch activity of magnetars appears to follow the same relation as for rotation-powered pulsars is the spin-down rate. Because there is almost no overlap between the spin frequencies and magnetic fields of magnetars, and those of the rotation-powered pulsars, the comparison is not possible for these parameters. Interestingly, however, it seems that the glitch activity of pulsars and magnetars does not appear to depend directly on their dipolar magnetic field strength. On the other hand, in the case of magnetars, some glitches are contemporaneous with X-ray bursts, which are thought to be powered by the decay of their strong magnetic fields \citep{2014ApJ...784...37D,2017ARA&A..55..261K}. Similarly, the largest glitches in the high magnetic field pulsars PSRs J1119$-$6127 and J1846$-$0258 were accompanied by changes in their emission properties \citep{2010ApJ...710.1710L,2011MNRAS.411.1917W,2016ApJ...829L..21A}. This possible connection between glitches and magnetospheric processes has lead to the idea that some glitches in high magnetic field neutron stars could have a different origin. Our results show, however, that the spin-down rate might be what determines the rate and size of these glitches, just as it does for ordinary pulsars.\\

We choose $|\dot{\nu}|$ as the best parameter to study $\dot{\nu}_{\rm{g}}$, because we can interpret our results in terms of simple physical concepts, and because this is the only parameter for which magnetars follow a similar tendency to rotation-powered pulsars. It is worth mentioning that we tested a broad range of combinations of the form $\nu^a\dot\nu$ with different values of $a$, finding that none of them give a substantially better correlation with $\dot\nu_{\rm{g}}$ than $|\dot\nu|$.\\

\begin{figure}[h!]
\centering
\includegraphics[width=90mm]{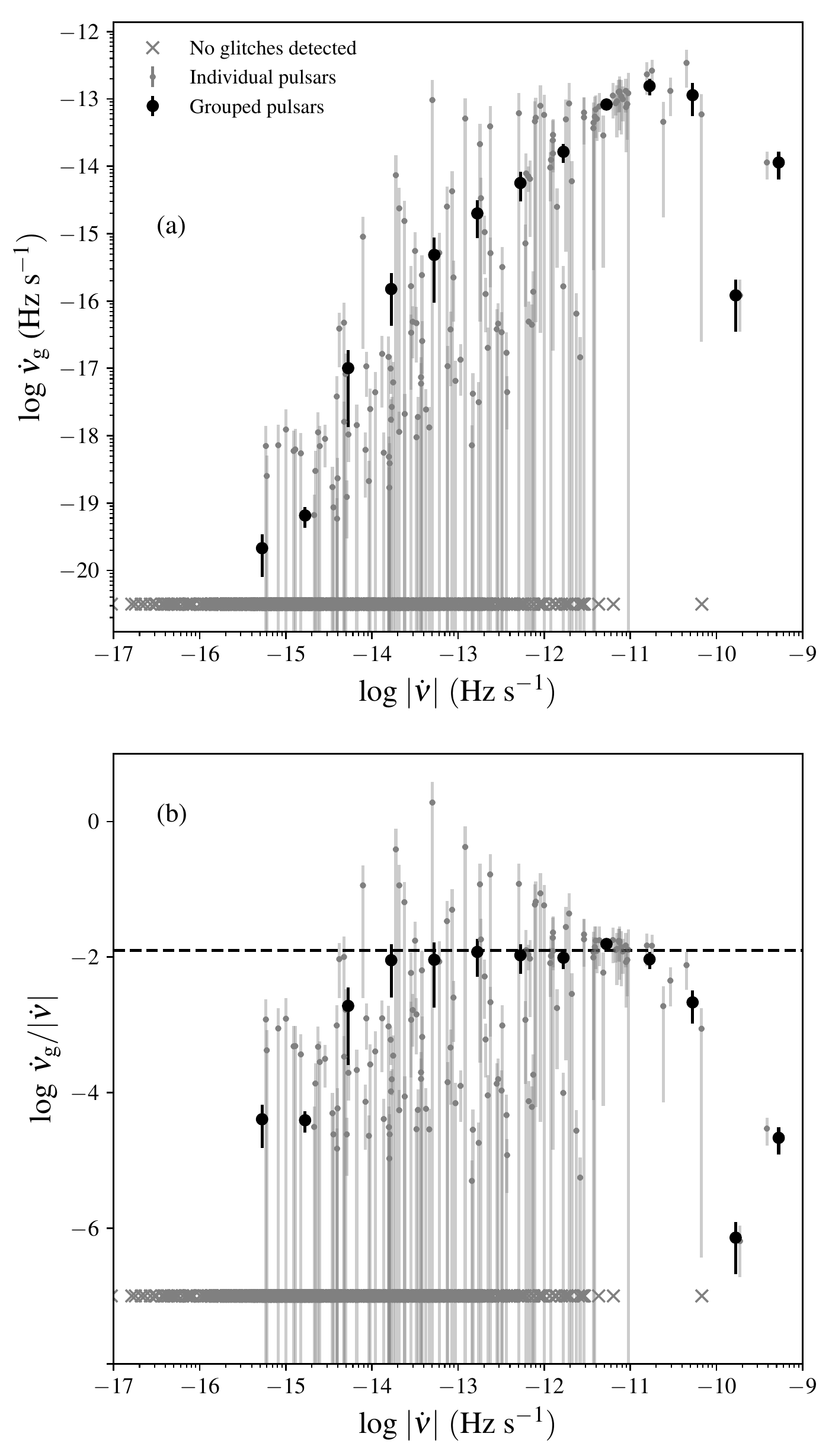}
\caption{Panel (a) shows $\log\,\dot{\nu}_{\rm{g}}$ versus $\log\,|\dot{\nu}|$. Panel (b) shows $\log\,(\dot{\nu}_{\rm{g}}/|\dot{\nu}|$) versus $\log\,|\dot{\nu}|$.  The horizontal line corresponds to the average ratio $\dot\nu_{\rm{g}}/|\dot\nu|= 0.012 \pm 0.001$, calculated over the bins with $-14 <\log\,|\dot{\nu}| < -10.5$. In both panels the crosses correspond to pulsars that have no glitches detected, whereas large black dots and small gray dots represent the bins (groups) and individual pulsars, respectively.} \label{fig4}
\end{figure}

Figure \ref{fig4} confirms the relation $\dot{\nu}_{\rm{g}} \propto |\dot{\nu}|$ already reported by \citet{2000MNRAS.315..534L} and \citet{2011MNRAS.414.1679E} for pulsars with $-14 < \log\,|\dot{\nu}| < -10.5$ (we always take the units of $|\dot{\nu}|$ as $\rm{Hz}$ $\rm{s^{-1}}$). The mean value of the ratio $\dot\nu_{\rm{g}}/|\dot\nu|$ for this range is $0.012 \pm 0.001$, corresponding to the horizontal line in Figure \ref{fig4}(b). This represents the fraction of the spin-down "recovered" by glitches, which can be interpreted as the fraction of the star's moment of inertia in a decoupled internal component \citep{1999PhRvL..83.3362L}. However, the linear trend between the glitch activity and the spin-down rate seems to fail towards the extremes, which we will explore further in \S \ref{large_glitches_section}. Table \ref{table1} shows additional information related to each bin in Figure \ref{fig4}.

\begin{table}[h!]
\centering
\caption{Statistics of glitches for pulsars binned by their spin-down rate.}
\begin{tabular}{@{}ccccccc@{}}
\toprule \toprule
$\#$ bin	& $\log\,|\dot{\nu}|$  & $\sum T_i$ & $N_{\rm{\ell}}$  & $N_{t}$ & $N_{\rm{pg}}$ & $N_{\rm{p}}$\\
& ($\rm{Hz}$ $\rm{s^{-1}}$) & ($\rm{yr}$) &  &  &\\
\midrule
1	& $-16.75$	&117	 & 0  &	0	&	0	& 7 \\
2	&$-16.25$	&430	 & 0  & 	0	&	0	&	25\\
3	&$-15.75$	&1233& 0  & 	0	&	0	&	70\\
4	&$-15.25$	&2478& 0  & 	3	&	3	&	139\\
5	&$-14.75$	&2675& 0  	& 	11	&	8	&	142\\
6	&$-14.25$	&1973& 0  	& 	25	&	16	&	105\\
7	&$-13.75$	&2083& 0  	&  	35	&	20	&	113\\
8	&$-13.25$	&1706& 1  	& 	29	&	18	&	105\\
9	&$-12.75$	&1312& 3  	&	26	&	14	&	81\\
10	&$-12.25$	&745& 4  & 	38	&	15	&	48\\
11	&$-11.75$	&493&  8   	&	74	&	15	&	33\\
12	&$-11.25$	&357&  37  	&	78	&	18	&	20\\
13	&$-10.75$	&66& 	13 	  &	19	&	5	&	5\\
14	&$-10.25$	&44& 4 & 	 8		&	2	&	3\\
15	&$-9.75$	&16 		& 0 & 	2	&	1	&	1\\
16	&$-9.25$	&46 		& 0 &  25	&	1	&	1\\
\hline
\end{tabular}
\tablefoot{The first column is the bin number. The second and third columns correspond to $\log\,|\dot{\nu}|$ for the group of pulsars in each bin (the central value of each logarithmic interval), and the sum of the observation time of all pulsars in that bin. The next two columns contain the number of large glitches and the total number of glitches, respectively. The last two columns correspond to the number of pulsars with glitches, and the total number of pulsars in each bin, respectively.}\label{table1}
\end{table} 

\subsection{A common trend in the activity of nearly all pulsars} \label{large_glitches_section}

\begin{figure}[h!]
\centering
\includegraphics[width=90mm]{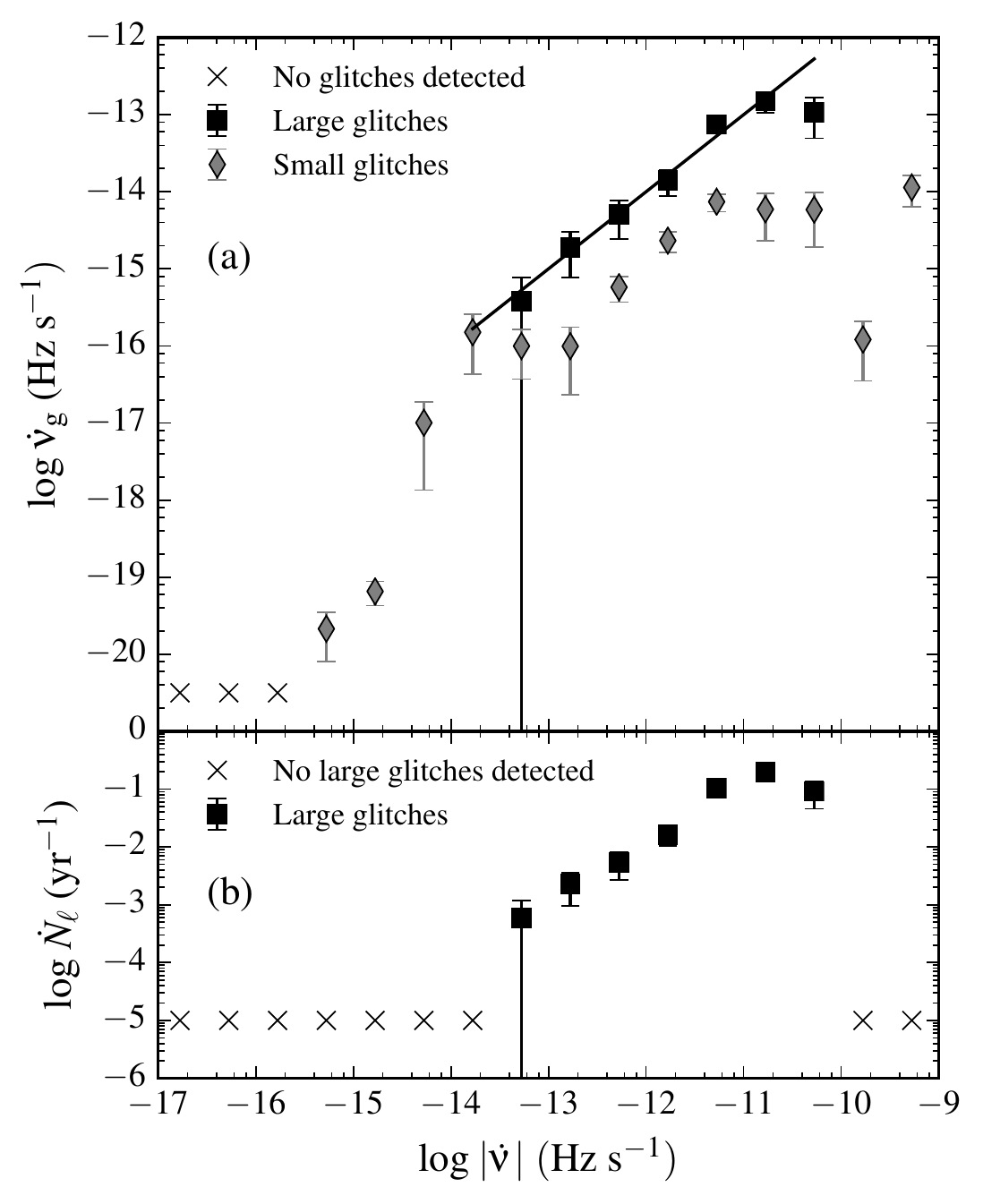}
\caption{Panel (a) shows $\log\,\dot{\nu}_{\rm{g}}$ versus $\log\,|\dot{\nu}|$. The black squares and gray diamonds represent the glitch activity separately for large and small glitches, respectively. In both cases, $\dot{\nu}_{\rm{g}}$ was calculated for the same bins (groups) of pulsars, respectively. The straight line shows the linear relation $\dot\nu_{\rm{g}}=0.01|\dot\nu|$. The crosses denote bins with no detected glitches. Panel (b) shows $\log\,\dot{N}_{\ell}$ versus $\log\,|\dot{\nu}|$. In panel (b) the black squares and the crosses represent the bins or groups of pulsars with large glitches and no large glitches detected, respectively.}\label{fig5}
\end{figure}

Motivated by the results in \S \ref{s2} showing that the largest glitches separate from the rest, we re-computed the glitch activity separately for large glitches ($\Delta \nu \geq 10\, \rm{\mu\, Hz}$) and for small glitches (the remainder), showing the results in Figure \ref{fig5}. We observe that large glitches determine the linear relation between $\dot{\nu}_{\rm{g}}$ and $|\dot{\nu}|$ in the range $-13.5< \log\,|\dot{\nu}| < -10.5$, and calculate the ratio $\dot{\nu}_{\rm{g}}/|\dot{\nu}| = 0.010 \pm 0.001$. This value is consistent with the value obtained in the previous section, when considering all glitches and including bin 7,  which has no large glitches detected. Accordingly, we use the linear relation $\dot\nu_{\rm{g}} = 0.01|\dot\nu|$ as a reference throughout the paper. The activity due to small glitches follows a roughly similar, though noisier, increasing trend with $|\dot\nu|$.\\

There are no large glitches detected in pulsars with the smallest and the largest spin-down rates ($\log\,|\dot{\nu}| < -13.5$ and $\log\,|\dot{\nu}| > -10$). The 14$^{th}$ bin ($\log\,|\dot{\nu}| = -10.25$, Table \ref{table1}) fails to follow the linear trend, despite having four large glitches with average size of $37\, \rm{\mu\, Hz}$. For the accumulated observing time in this bin, approximately 30 additional large glitches of 20 $\rm{\mu\, Hz}$ are required to reach the activity value  predicted  by the linear relation. On the other side, the 7$^{th}$ bin ($\log\,|\dot{\nu}| = -13.75$) follows the linear trend even though it has no large glitches detected. According to the linear relation and the accumulated observing time, only one large glitch of $10\, \rm{\mu\, Hz}$ is necessary to obtain the value predicted by the linear trend, which is very close to the current activity value. However, instead of one large glitch, there are smaller glitches that account for 9.78 $\rm{\mu\, Hz}$, making the activity consistent with the linear relationship.\\

Since the proportionality between the glitch activity and the spin-down rate is dominated by large glitches, and these have a very narrow size distribution, we expect that the rate of large glitches, $\dot{N}_{\rm{\ell}}$, will also be proportional to $|\dot{\nu}|$ (Fig. \ref{fig5}b). Because the number of large glitches is expected to follow a Poisson distribution, the expected dispersion in the rate of large glitches $\dot{N}_{\rm{\ell}}$ can be estimated in a more reliable way than that in the glitch activity $\dot\nu_{\rm{g}}$. This allows us to test in a statistically meaningful way whether or not the identified trend applies to all pulsars.\\

\begin{figure}[h!]
\centering
\includegraphics[width=88mm]{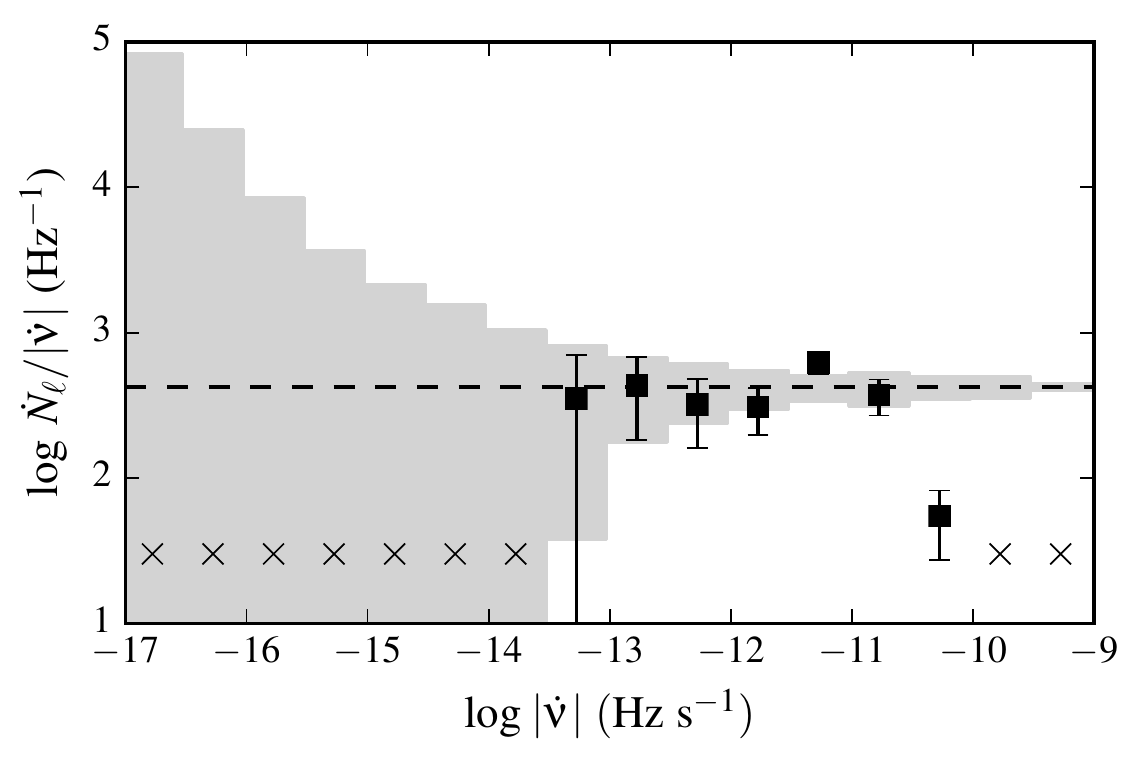}
\caption{$\log\, \, \dot{N}_{\rm{\ell}}/|\dot{\nu}|$ versus $\log\,|\dot{\nu}|$, where $\dot N_{\ell}$ is the number of large glitches per unit time. The horizontal line corresponds to the logarithm of the mean value $\langle\dot{N}_{\rm{\ell}}/|\dot{\nu}|\rangle = (4.2\pm 0.5)\times 10^2\, \rm{Hz^{-1}}$, calculated over the bins with $-13.5 < \log\, |\dot{\nu}| < -10.5$. The shaded region indicates the expected dispersion around this average value, based on a Poisson distribution of the number of large glitches and the available observing time spans. The black squares represent the observed values of the ratio $\dot{N}_{\rm{\ell}}/|\dot{\nu}|$ for bins (groups) of pulsars. The crosses denote bins with no large glitches detected.} \label{fig6}
\end{figure}

Figure \ref{fig6} confirms that $\dot{N}_{\rm{\ell}}/|\dot{\nu}|$ is approximately constant and its mean value is $(4.2 \pm 0.5)\times 10^{2}$ $\rm{Hz^{-1}}$ (which we calculated considering only the $|\dot{\nu}|$ bins for pulsars with $-13.5 < \log\,|\dot{\nu}| < -10.5$). We observe that except for the three bins with the largest spin-down rate, all others are consistent with this trend. The non-detection of large glitches in the region of small $|\dot\nu|$ is consistent with the small expected rate and the finite monitoring time, as illustrated by the shaded area in Figure \ref{fig6}. Based on this relation, the expected number of large glitches ($N^{\rm{exp}}_{\rm{\ell}}$) for the three bins with the highest $|\dot\nu|$ ($\log |\dot{\nu}|> -10.5$) is $30\pm 5$, $35\pm 5$, and $325\pm 18$, respectively. This strongly contradicts the only four large glitches detected in bin 14 and the absence of large glitches in bins 15 and 16 (which contain only PSR B0540$-$69 and the Crab pulsar, respectively; see Table \ref{table1}). Thus, we can confidently rule out the linear relation between $\dot{\nu}_{\rm{g}}$ and $|\dot{\nu}|$ for the largest values of the latter variable, but it remains consistent for all $|\dot{\nu}| < 10^{-10.5}$ $\rm{Hz\, s^{-1}}$.\\

Next, we test whether the individual pulsars within each bin are also consistent with this trend. Since the number of large glitches for each pulsar is small (in most cases zero), the usual $\chi^2$ test is not applicable. Instead, we use Fisher's test \citep{fisher1925statistical}, based on the statistic 
\begin{equation} 
X^2_{2k} = -2\sum_{i=1}^k \ln\, p_i\text{,} \label{fisher}
\end{equation}
with two-tailed $p$-values for each pulsar calculated as 
\begin{equation}
p_i = \min\{P(x \leq N^{\rm{obs}}_{\rm{\ell}})\, ,\, P(N^{\rm{obs}}_{\rm{\ell}}\geq x)\}\text{,}
\end{equation}
\noindent where $P(x \leq N^{\rm{obs}}_{\rm{\ell}})$ is the (Poisson) probability of obtaining a value $x$ smaller or equal to the actual observed value $N^{\rm{obs}}_{\rm{\ell}}$, based on the fixed ratio $\dot {N_{\ell}}/|\dot{\nu}|= (4.2\pm 0.5)\times 10^2\, \rm{Hz^{-1}}$ calculated above, and the observation time of the pulsar (and analogously for $P[N^{\rm{obs}}_{\rm{\ell}}\geq x])$.\\

If the null hypothesis is true, $p_i$ is uniformly distributed between 0 and 1, and therefore the Fisher statistic will follow a $\chi^2$ distribution with $2k$ degrees of freedom. On the other hand, when the individual p-values $p_i$ are very small, $X^2_{2k}$ will be large, leading to the rejection of the global null hypothesis.\\

Table \ref{fisher_results} shows the results of the test, where the global $p$-value for each bin was calculated as
\begin{equation}
p_{\rm{bin}} = P(\chi^2 \geq X^2_{2k}\, |\, X^2_{2k} \sim \chi^2_{2k})\text{,} \label{pbin}
\end{equation}
that is, the probability of obtaining a $\chi^2$ value at least as large as our observed $X^{2}_{2k}$, under the null hypothesis $X^2_{2k} \sim \chi^2_{2k}$. Based on the values obtained for $p_{\rm{bin}}$, the null hypothesis can be strongly ruled out for the last three bins, whereas we cannot rule out that all pulsars with $|\dot{\nu}| < 10^{-10.5}$ $\rm{Hz\, s^{-1}}$, even those without glitches, follow the identified trend.

\begin{table}[h!]
\centering
\caption{Results of Fisher's method} \label{fisher_results}
\begin{tabular}{@{}lll@{}}
\toprule \toprule
$\#$ bin & $X^2_{2k}\, $\text{ / dof} & $p_{\rm{bin}}$\\
\midrule
1 & $0.00005$ \text{ / 12} & $1.0$\\
2 & $0.0007$ \text{ / 50} & $1.0$\\
3 & $0.006$ \text{ / 140} & $1.0$\\
4 & $0.04$ \text{ / 276}  & $1.0$\\
5 & $0.12$ \text{ / 286} & $1.0$\\
6 & $0.30$ \text{ / 210} & $1.0$\\
7 & $0.96$ \text{ / 226} & $1.0$\\
8 & $13.29$ \text{ / 210} & $1.0$\\
9 & $23.73$ \text{ / 162} & $1.0$\\
10 & $25.23$ \text{ / 96} &$1.0$\\
11 & $32.13$ \text{ / 66} &$0.99$\\
12 & $40.42$ \text{ / 40} &$0.49$\\
13 & $18.09$ \text{ / 10} &$0.07$\\
14 & $57.24$ \text{ / 6} &$2\times 10^{-10}$\\
15 & $79.30$ \text{ / 2} &$3\times 10^{-18}$\\
16 & $475.21$ \text{ / 2} &$3\times 10^{-104}$\\
\bottomrule
\end{tabular}
\tablefoot{The first column denotes the bin number, whereas the second and third columns contain the Fisher statistic (Eq. \ref{fisher}) and the $p$-value defined in Eq.\ref{pbin}.}
\end{table}

\section{Discussion} \label{disc}
We have shown that all pulsars with $|\dot\nu|< 10^{-10.5}\, \rm{Hz\, s^{-1}}$ and magnetars are consistent with a single trend, dominated by large glitches, in which the glitch activity $\dot{\nu}_{\rm{g}}$ is equal to $0.01|\dot{\nu}|$. The large collection of pulsars with no detected glitches is also consistent with this trend. For instance, the predicted rate of large glitches for pulsars with $\log\, |\dot{\nu}| = -14.25$ (bin 6) is one large glitch every $\sim 10^4$ yr, whereas the accumulated observing time in this bin is only 1973 yr (Table \ref{table1}), and this mismatch becomes even more extreme for the bins with lower $|\dot\nu|$. This means that there are no reasons to reject the idea that every neutron star will eventually experience a large glitch and will, in the long-term, follow the above relationship. On the other hand, we cannot rule out the possibility that, for example, the glitch mechanism could fail to produce large glitches in the pulsars with the smallest spin-down rates, or produce substantially more of them than predicted from the linear relation.\\

A similar amount of superfluid neutrons inside all neutron stars could be responsible for the common glitch activity levels observed \citep{1999PhRvL..83.3362L}. However, despite this uniformity in the collective behavior, it appears that the glitch mechanism possesses an intrinsic bi-modality (Fig. \ref{fig2}). It could be that the mechanism results in different glitch sizes and rates depending on where in the star the glitch happens \citep{2012MNRAS.420..658H,2014MNRAS.438L..16H}. An alternative interpretation is that there are two or more different glitch mechanisms in action, giving rise to different size distributions.\\

The cumulative distributions of glitch sizes for five individual pulsars, presented in Figure 16 of \citet{2011MNRAS.414.1679E}, suggests that this bi-modality could be extended to define at least two types of pulsars: those with large glitches and those with only small glitches; but we find no evidence for this in the glitch activity data.

\subsection{The glitch behavior of the most rapidly evolving pulsars}
\begin{figure}[h!]
\centering
\includegraphics[width=88mm]{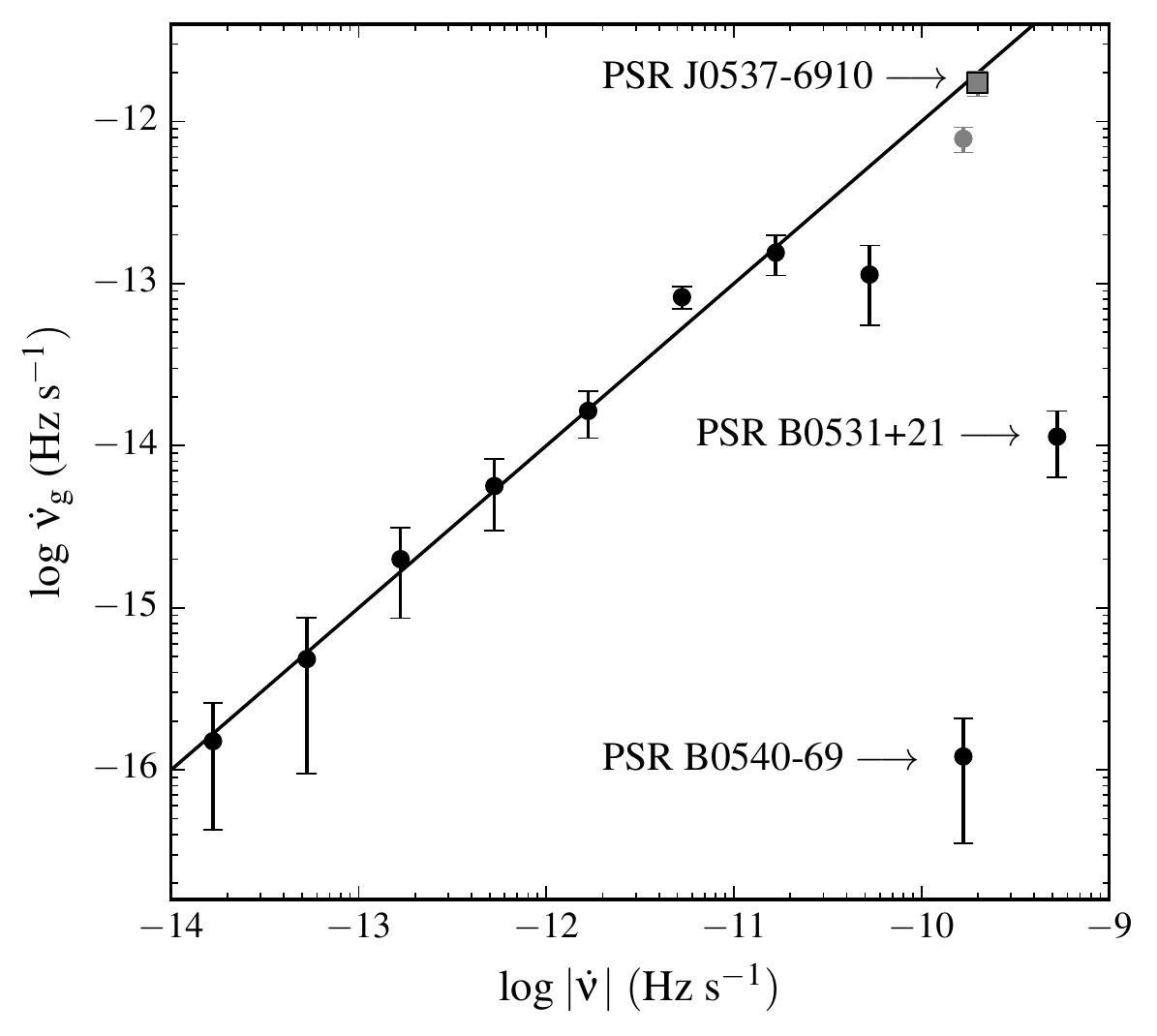}
\caption{A zoom-in to the zone $\dot{\nu}_{\rm{g}}=0.01|\dot{\nu}|$ plus the last three bins, with the highest $|\dot{\nu}|$ values. The straight line shows the linear relation $\dot\nu_{\rm{g}}=0.01|\dot\nu|$. The glitch activity of PSR J0537$-$6190 was plotted using a gray square to show that this young pulsar follows the general trend. The activity obtained when including this pulsar in the appropriate bin is represented by the gray dot.} \label{j0537}
\end{figure}

The glitch activity for the last two bins was calculated using only one pulsar per bin (see Table \ref{table1}) and the values obtained do not follow the tendency defined by the rest of the population. These are the pulsars with the largest spin-down rates and correspond to PSR B0540$-$69 (bin 15) and the Crab pulsar (PSR B0531$+$21, bin 16). PSR B0540$-$69 has been observed for $\sim 15.8$ $\rm{yr}$ \citep{2015ApJ...812...95F} and its activity is especially low, compared to all pulsars with high spin-down rates (see Fig. \ref{fig4}). In the same bin it is possible to include the X-ray pulsar PSR J0537$-$6910, which was not considered in our sample since it does not belong to any of the monitoring programs taken into account for our database, and including it would have artificially biased the sample towards the presence of glitches. It has been observed for $13$ $\rm{yr}$, and $45$ glitches have been detected \citep{2004ApJ...603..682M,2006ApJ...652.1531M,2017arXiv170809459A,2017arXiv170808876F}, 30 of which are large according to our classification in \S \ref{s2}, making it consistent with the linear trend identified above. If we included this pulsar in our dataset, the glitch activity in the associated bin would be only slightly lower than predicted by the relation $\dot{\nu}_{\rm{g}} =0.01|\dot{\nu}|$ (see Fig. \ref{j0537}).\\

\citet{2015ApJ...807L..27M} reported a large increase in the spin-down rate of PSR B0540$-$69 that remained for $< 3$ years. Based on their data, they placed a limit of $\Delta \nu < 12$  $\rm{\mu\, Hz}$ for a hypothetical glitch responsible for this increase. With a glitch of the maximum allowed size, the activity of PSR B0540$-$69 would be $2(2) \times 10^{-14}$ $\rm{Hz}$ $\rm{s^{-1}}$, very similar to the activity of the Crab pulsar $1.1(5)\times 10^{-14}$ $\rm{Hz}$ $\rm{s^{-1}}$, but still low if the pulsar was to follow the relationship above.\\

One possible explanation for the discrepancy in the glitch activity of these pulsars could be their age, as proposed by \citet{1996ApJ...459..706A} to explain the differences between the Crab and Vela pulsars. Indeed, PSR B0540$-$69 and the Crab pulsar are among the youngest pulsars known, with supernova remnant ages equal to $1000^{+600}_{-240}$ and 962 yr, respectively \citep{2010ApJ...710..948P,2012NatPh...8..787H}. However, other similarly young pulsars, such as PSRs J1119$-$6127 and J1846$-$0258 (in bins 13 and 14, respectively), but with lower spin-down rates, have experienced large glitches during their monitored rotations and exhibit glitch activities in closer agreement with the main trend. We conclude that it is possible that the glitch mechanism might not be able to operate normally when the spin-down rate is too high. Perhaps, once PSR B0540$-$69 and the Crab pulsar have evolved, and their spin-down rates have decreased, their glitches will have settled into the trend followed by the rest of the population.

\section{Conclusions} \label{summary}
In this paper we present a statistical study of glitches in pulsars and magnetars, using a large database of pulsars whose selection was independent of their glitch properties. The main conclusions are the following:
\begin{itemize}
\item[1)]The glitch size distribution is at least bimodal, with two well-defined classes: a broad distribution of small glitches and a narrow one with large glitches peaked at around $20\, \rm{\mu\, Hz}$. However, there is no evidence in the glitch activity data for the existence of two classes of pulsars.\\

\item[2)]For pulsars and magnetars with $10^{-13.5} < |\dot{\nu}| < 10^{-10.5}$ $\rm{Hz\, s^{-1}}$, the glitch activity $\dot{\nu}_{\rm{g}}$ is directly proportional to the spin-down rate $|\dot{\nu}|$, following  $\dot\nu_{\rm{g}} = (0.010 \pm 0.001)\, |\dot\nu|$. This relationship is dominated by large glitches. For all pulsars in this group, the rate of large glitches is consistent with the proportionality relation $\dot{N}_{\rm{\ell}} = (4.2 \pm 0.5)\times 10^{2}\, \rm{Hz^{-1}}\, |\dot{\nu}|$. This is also consistent for the pulsars with $|\dot{\nu}| < 10^{-13.5}$, which have not been observed long enough to detect large glitches. Thus, we showed that the glitch activity of every pulsar with $|\dot{\nu}| < 10^{-10.5}$, including those with no glitches, is statistically consistent with the above relationships.\\

\item[3)]The activity due to small glitches also increases with $|\dot{\nu}|$. We did not present an explicit relation between the activity of small glitches and $|\dot{\nu}|$ because it depends strongly on the choice of the cutoff value between small and large glitches, and on detectability issues.\\

\item[4)]Pulsars with $|\dot{\nu}| > 10^{-10.5}$ $\rm{Hz\, s^{-1}}$ present an intrinsically different behavior, with a much lower rate of large glitches.
\end{itemize} 

Future studies, based on the analysis of glitches in individual pulsars, should provide information helping clarify whether there are two or more glitch mechanisms giving rise to the observed multimodal glitch size distribution. For this, well designed, long-term monitoring campaigns are needed, in which cadence and sensitivity are combined to ensure the detection of small glitches, thereby improving the completeness of the current samples.

\begin{acknowledgements}
We thank Danai Antonopoulou for useful comments. This work was funded by ALMA-CONICYT Astronomy/PCI Project 31140029, FONDECYT Regular Projects 1150411 and 1171421, CONICYT Basal Funding Grant PFB-06, CONICYT grant PIA ACT1405, and a SOCHIAS travel grant through ALMA/Conicyt Project $\#$31150039. We thank the observers at Jodrell Bank Observatory. Pulsar research at Jodrell Bank Centre for Astrophysics (JBCA) is supported by a consolidated grant from the UK Science and Technology Facilities Council (STFC).
\end{acknowledgements}

\bibliographystyle{aa}
\bibliography{references}
\begin{appendix}
\section{Model selection: Akaike's information criterion applied to the glitch size distribution} \label{akai}
Given a model $M$ that includes $k$ parameters $\theta_p$, $p = 1, . . . , k$, abbreviated as the vector $\mathbf{\theta}$ with components $\theta_{p}$, we can compute the likelihood $p(x_i|\mathbf{\theta})$ of observing any given datum $x_i$. If we assume that our data $\{x_i\}$ are all independent of each other, the likelihood of the whole dataset is given by
\begin{equation}
L = p(\{x_i\}|\mathbf{\theta}) = \prod_{i} p(x_i|M(\mathbf{\theta})) \text{.}
\end{equation}
Once the likelihood for a set of models is known, the model with the largest value provides the best description of the dataset. However, this is not necessarily the best model overall when models have different numbers of free parameters. Akaike's information criterion (AIC) is defined as 
\begin{equation}
AIC = -2\ln L + 2k\text{,}
\end{equation}
\citep{1974ITAC...19..716A}, where $L$ is the likelihood of the dataset given the model, and $k$ is the number of free parameters of the model. According to the AIC, the best model is the one with the smallest AIC value. The AIC penalizes for the addition of parameters, and therefore selects a model that fits well and has a minimum number of parameters. We note that the AIC value by itself has no meaning, and only has significance when compared to the values for a set of alternative models.\\

In order to compare the models, we need two measures: the relative AIC and Akaike weights. The relative AIC is the difference of a given candidate model $m$ in respect to the model with the minimum AIC value 
\begin{equation}
\Delta_m = (AIC)_m - (AIC)_{\rm{min}}\text{,}
\end{equation}
where $(AIC)_m$ is the AIC value for the model $m$, and $(AIC)_{\rm{min}}$ is the AIC value of the best candidate model. Akaike weights $w_m$ represent the probability that the model $m$ is the best model (in the AIC sense, that it minimizes the loss of information), given the data and the whole set of candidate models:
\begin{equation}
w_m = \dfrac{\exp(-\Delta_m/2)}{\sum_k \exp(-\Delta_k/2)}\text{,} \label{w_i}
\end{equation}
so that $\sum_m w_m = 1$.\\

In order to test if the glitch size distribution is multimodal, we modeled it as a sum of $M$ Gaussians, writing the likelihood of a datum $x_i = \log\Delta\nu_i$ as
\begin{equation}
p(x_i|\boldmath{\theta}) = \sum_{j=1}^{M}\alpha_j \dfrac{1}{\sqrt{2\pi}\sigma_j}e^{-(x_i - \mu_j)^2/2\sigma_j^2}\text{,}
\end{equation}
\noindent where $\boldmath{\theta}$ denotes the vector of parameters that need to be estimated; it includes normalization factors for each Gaussian, $\alpha_j$ (with the constraint that $\sum_j\alpha_j=1$), and its mean $\mu_j$ and dispersion $\sigma_j$. Using the Expectation-Maximization algorithm described in \citet{Ivezic:2014:SDM:2578955}, we obtained the parameters that maximize the log-likelihood of the whole dataset $\{x_i\}$, for models with one, two, three, and four Gaussians. In order to decide the model that gives the best description of the data, we applied the AIC and calculated the Akaike weights $w_i$ for each model according to Equation \ref{w_i}. Table \ref{akairesults} summarizes the results.
\begin{table}[h!]
\centering
\caption{Results of the Expectation - Maximization algorithm and the AIC applied to the glitch size distribution modeled as a sum of Gaussians.}
\begin{tabular}{@{}lcccc@{}}
\toprule \toprule
Parameters & $1$-G & $2$-G & $3$-G & $4$-G\\
\midrule
$\alpha_1$ & 1 & 0.83 & 0.36 & 0.27 \\ 
$\mu_1$ & -1.02 & -1.51 & -2.62 & -2.72  \\ 
$\sigma_1$ & 1.65 & 1.37 & 0.79 & 0.73 \vspace{1mm}  \\ 
$\alpha_2$ & ... & 0.17 & 0.47 & 0.26  \\ 
$\mu_2$ & ... & 1.37 & -0.61 & -1.48 \\ 
$\sigma_2$ & ... & 0.13 & 0.13 & 1.17 \vspace{1mm}  \\ 
$\alpha_3$ & ... & ... & 0.17 & 0.30  \\ 
$\mu_3$ & ... & ... & 1.38  & -0.41 \\ 
$\sigma_3$ & ... & ... & 0.13 & 1.01 \vspace{1mm}  \\ 
$\alpha_4$ & ... & ... & ... & 0.17  \\ 
$\mu_4$ & ... & ... & ... & 1.38 \\ 
$\sigma_4$ & ... & ... & ... & 0.13 \vspace{1mm}  \\ 
$AIC$ & 1437.81 & 1309.42 & 1297.17 & 1303.48 \\ 
$\Delta_m$ & 140.64 & 12.25 & 0 & 6.31 \\ 
$w_m$ & $10^{-31}$ & 0.002  & 0.957 & 0.040 \\ 
\bottomrule
\end{tabular}
\tablefoot{The label $m$-G in each column represents the model $m$ (i.e., a sum of $m$ gaussians). $\{\alpha_j,\, \mu_j,\, \sigma_j\}$ correspond to the weight, mean, and the standard deviation of the $j$-th component of the sum of gaussians, respectively. $\{AIC,\, \Delta_m,\, w_m\}$ denote the AIC value, the relative AIC respect to the model with the minimum AIC value, and the Akaike weights of the model $m$, respectively.}\label{akairesults}
\end{table}
The Akaike weight associated to the model with three components is $w_3 = 0.957$, that is, it has a $95.7\, \%$ probability of being the best one among the candidate models considered.

\section{Error estimation of the glitch activity} \label{errors}
In order to estimate the error in $\dot{\nu}_{\rm{g}}$, we assume that detections of glitches of any given size follow Poisson statistics. If we detected $N$ glitches with a magnitude equal to $\Delta \nu$, in an observation span $T$, we can write the glitch activity as
\begin{equation}
\dot{\nu}_{\rm{g}} = \dfrac{N\Delta \nu}{T}\text{,}
\end{equation}
and its variance as
\begin{equation}
\text{Var}\left[\dot{\nu}_{\rm{g}}\right] = \left(\dfrac{\Delta \nu}{T}\right)^2\text{Var}\left[N\right] = \left(\dfrac{\Delta \nu}{T}\right)^2N\text{.}
\end{equation}
Thus, the error is estimated as
\begin{equation}
\delta \dot{\nu}_{\rm{g}} = \sqrt{\text{Var}\left[\dot{\nu}_{\rm{g}}\right]} = \dfrac{\sqrt{N}\Delta \nu}{T} = \dfrac{\dot{\nu}_{\rm{g}}}{\sqrt{N}}\text{.}
\end{equation}
In order to generalize this expression for glitches of different magnitude, we assume that the $N$ glitches are distributed in $M$ groups, each one with $N_k$ glitches of the same size $\Delta \nu_k$. In this case, the glitch activity is written as 
\begin{equation}
\dot{\nu}_{\rm{g}} = \sum_{k=1}^{M} N_k \left(\frac{\Delta \nu_k}{T}\right) \text{,}
\end{equation}
and the variance
\begin{equation}
\text{Var}\left[\dot{\nu}_{\rm{g}}\right] = \sum_k\left(\Delta\nu_k\over T\right)^2 \text{Var}\left[N_k\right] = \sum_k \dfrac{N_k \Delta \nu_{k}^2}{T^2} \text{.} \label{vg}
\end{equation}
If we consider that all glitches have different sizes, the last sum in the Equation \ref{vg} can be rewritten as
\begin{equation}
\dfrac{1}{T^2} \sum_j \Delta \nu_{j}^2\text{, } \Longrightarrow \delta \dot{\nu}_{\rm{g}} = \dfrac{1}{T} \sqrt{\sum_j \Delta \nu_{j}^2}\text{,}
\end{equation}
where the sum runs over all the glitches.
\end{appendix}
\end{document}